\documentclass{ieeeaccess}
\usepackage{cite}
\usepackage{amsmath,amssymb,amsfonts}
\usepackage{algorithmic}
\usepackage{graphicx}
\usepackage{textcomp}

\usepackage{bm}
\makeatletter
\AtBeginDocument{\DeclareMathVersion{bold}
\SetSymbolFont{operators}{bold}{T1}{times}{b}{n}
\SetSymbolFont{NewLetters}{bold}{T1}{times}{b}{it}
\SetMathAlphabet{\mathrm}{bold}{T1}{times}{b}{n}
\SetMathAlphabet{\mathit}{bold}{T1}{times}{b}{it}
\SetMathAlphabet{\mathbf}{bold}{T1}{times}{b}{n}
\SetMathAlphabet{\mathtt}{bold}{OT1}{pcr}{b}{n}
\SetSymbolFont{symbols}{bold}{OMS}{cmsy}{b}{n}
\renewcommand\boldmath{\@nomath\boldmath\mathversion{bold}}}
\makeatother

\def\BibTeX{{\rm B\kern-.05em{\sc i\kern-.025em b}\kern-.08em
    T\kern-.1667em\lower.7ex\hbox{E}\kern-.125emX}}

\usepackage{algorithm}
\usepackage{algorithmic}
\usepackage{tabularx}

\usepackage{multirow}
\usepackage{tabularx}
\usepackage{float}
\usepackage{algorithm}
\usepackage{algorithmic}
\usepackage{amsmath} 
\usepackage{scalerel}

\usepackage{makecell}
\usepackage[hidelinks]{hyperref}
\usepackage{fancyhdr}
\fancypagestyle{preprintstyle}{
    \fancyhf{} 
    \fancyhead[L]{\textit{Preprint}} 
    \fancyfoot[C]{\thepage}          

}

\providecommand{\history}[1]{}
\providecommand{\doi}[1]{}
\providecommand{\markboth}[2]{}
\providecommand{\corresp}[1]{}

\makeatletter
\def\@history{}
\def\@doi{}
\def\@markboth#1#2{}
\def\@corresp{}
\makeatother
\begin{document}
\title{Compressed Learning for Nanosurface Deficiency Recognition Using Angle-resolved Scatterometry Data}
\author{
\uppercase{Mehdi Abdollahpour}\authorrefmark{1},
\uppercase{Carsten  Bockelmann}\authorrefmark{1}, 
\uppercase{Tajim Md Hasibur Rahman}\authorrefmark{2}, 
\uppercase{Armin Dekorsy} \authorrefmark{1},
and \uppercase{Andreas Fischer}, \authorrefmark{2}}

\address[1]{Department of Communications Engineering, University of Bremen, 28359 Bremen, Germany (e-mail: abdollahpour@ant.uni-bremen.de, bockelmann@ant.uni-bremen.de, dekorsy@ant.uni-bremen.de)}
\address[2]{Bremen Institute for Metrology, Automation and Quality Science, University of Bremen, 28359 Bremen, Germany (e-mail: t.rahman@bimaq.de, andreas.fischer@bimaq.de)}



\begin{abstract}
Nanoscale manufacturing requires high-precision surface inspection to guarantee the quality of the produced nanostructures. 
For production environments, angle-resolved scatterometry offers a non-invasive and in-line compatible alternative to traditional surface inspection methods, such as scanning electron microscopy. However, angle-resolved scatterometry currently suffers from long data acquisition time.
Our study addresses the issue of slow data acquisition by proposing a compressed learning framework for the accurate recognition of nanosurface deficiencies using angle-resolved scatterometry data. 
The framework uses the particle swarm optimization algorithm with a sampling scheme customized for scattering patterns. This combination allows the identification of optimal sampling points in scatterometry data that maximize the detection accuracy of five different levels of deficiency in ZnO nanosurfaces.
The proposed method significantly reduces the amount of sampled data while maintaining a high accuracy in deficiency detection, even in noisy environments.
Notably, by sampling only 1\% of the data, the method achieves an accuracy of over 86\%, which further improves to 94\% when the sampling rate is increased to 6\%.  These results demonstrate a favorable balance between data reduction and classification performance. 
The obtained results also show that the compressed learning framework effectively identifies critical sampling areas.
\end{abstract}



\maketitle

\thispagestyle{preprintstyle} 
\pagestyle{preprintstyle}      


\section{Introduction}

Nanoscale manufacturing has become an important part of the production lines for various fields of application by making enhancements in transistors \cite{sze2021physics}, energy efficiency of solar cells \cite{zheng2016photonic}, and diagnostics in medicine \cite{ferrari2005cancer}. 
The manufactured nanostructures need to be monitored for quality and detection of deficiencies \cite{madsen2015fast, alexe2018model}. 
Scanning electron microscopy (SEM) and atomic force microscopy (AFM) are the available techniques that can be used to analyze nanosurfaces. However, AFM requires direct contact with the nanosurface, which may disturb delicate samples, and SEM often requires conductive coating, making it partially destructive  \cite{goldstein2017scanning, fukuma2012atomic}. These limitations make them  unsuitable for use in a manufacturing environment where rapid, non-destructive, and real-time monitoring of nanosurfaces is required \cite{madsen2016scatterometry}. 

Scatterometry is an alternative technique to SEM and AFM for surface inspection and dimensional metrology \cite{orji2018metrology}.
It is a non-contact technique that analyzes the interaction of light with a nanostructured surface. In scatterometry, a coherent light source is directed onto the surface, and the scattered light is measured using a light-detecting sensor to extract structural information \cite{whitworth2020real}. 
Scatterometry is widely used in the semiconductor industry for the inspection and metrology of manufactured computer chips. It provides a non-destructive and high-resolution method for critical dimension measurements and process control  \cite{den2017scatterometry}.
It is also shown to be applicable in monitoring crystal growth processes \cite{heurlin2015situ} and injection-molded plastic \cite{zalkovskij2015smart}.

A specific type of scatterometry is called angle-resolved scatterometry (ARS). In ARS, a light detector moves around the nanostructured sample to measure scattered light from all reflection angles across the surrounding spherical field.
ARS provides comprehensive angular characterization of the surface by capturing a complete image of scatterometric data that delivers detailed information about the surface characteristics \cite{jovst2014camera}.
The provided comprehensive information from angle-resolved data can help to  analyze nanostructures for surface roughness \cite{kato2010effect, amra1993multiwavelength}, thin film characterization \cite{zhang2023situ} and 3D features metrology \cite{charley20113d}.

ARS has been shown to be useful for the detection of local defects in nanostructured surfaces \cite{lonardo1991surface}.
In a study, Zimmerman et al.\cite{zimmermann2012process} designed a framework for the detection of defective sub-100 nm structures in zinc oxide (ZnO) nanosurfaces.
They used discrete dipole approximation (DDA) \cite{yurkin2011discrete} to generate computer-simulated data for further evaluation of the scattered light. In addition, an ARS device was built for the comparison of the simulated data with the real data.
However, Zimmerman's study did not carry out a quantitative analysis on the detection of deficiencies in surfaces.
In addition, the ARS device should scan the encompassing sphere of the nanosurface to record the scattered light from all reflection angles. This task is normally carried out using a single moving light sensor, which makes this process significantly slow.

Similar to Zimmerman's study, we will also study the computer-simulated ARS data acquired from the ZnO nanosurface. ZnO is a semiconductor known for its excellent optical transparency, high electron mobility, and strong room-temperature luminescence. These properties make it highly suitable for applications in optoelectronics, sensors, and surface coatings \cite{ozgur2005comprehensive}. In these applications, the performance of ZnO-based components is highly sensitive to surface quality, as surface defects or irregularities can negatively impact optical and electrical behavior \cite{mutar2025fabrication, manabeng2022review}. Therefore, surface inspection plays an important role in maintaining the functionality of the final product.

Additionally, our study addresses the mentioned challenges of quantitative analysis of the nanosurface deficiency levels and faster data acquisition in ARS by utilizing compressed learning (CL). CL combines compressed sensing \cite{donoho2006compressed, candes2008introduction} with machine learning in an integrated framework.
In CL, the objective is shifted from the reconstruction of the data, which was the case in compressed sensing, to a higher-level machine learning task like classification. Therefore, instead of the reconstruction of the data using the compressed measurements and then carrying out a machine learning task, the data in the measurement domain can be investigated directly using the machine learning algorithms \cite{adler2016compressed}.

A related study presents a unified framework that connects sparse signal modeling and supervised learning within the context of compressed learning \cite{calderbank2009compressed}.  They show that it is possible to carry out learning tasks directly in the compressed domain without recovering the original data.
The work establishes theoretical guarantees and provides practical results showing that compressed learning can achieve comparable or even superior performance to traditional learning methods applied to uncompressed data, especially in settings where data acquisition and storage are limited.


CL has been applied across various domains as a practical tool for efficient data analysis \cite{zisselman2018compressed}. In a study, Lohit et al. \cite{lohit2015reconstruction} demonstrated that high-level visual inference tasks, such as face recognition, can be performed using a compressive camera. Their approach combined smashed filters \cite{davenport2007smashed} with support vector machines to allow direct inference from compressed measurements.

CL has also been utilized in biological data analysis. Zhang et al. \cite{zhang2011adaptive} applied CL techniques to reduce the dimensionality of protein sequence feature vectors for the prediction of protein-protein interactions. This reduction in dimensionality helped lower computational complexity while preserving the predictive performance of the model. These studies illustrate the adaptability of CL in handling high-dimensional data across different scientific and engineering fields.


We modified the traditional CL framework to make significant improvements to ARS sampling and analysis. 
Traditional compressed learning builds on compressed sensing by using a measurement matrix \( \mathbf{\Phi} \in \mathbb{R}^{m \times p} \), where \( m \ll p \), to obtain compressed measurements \( \mathbf{y} = \mathbf{\Phi x} \) from the original signal $\mathbf{x}$.
Instead of reconstructing \( \mathbf{x} \), learning tasks such as classification are performed directly on \( \mathbf{y} \). Because sparsity is not a prerequisite in compressed learning, the measurement matrix \( \mathbf{\Phi} \) can be learned or optimized for better performance.
While this method of sampling can be implemented in an analog domain signal sampler, its applicability in position-based sampling like ARS is limited.



Our approach adapts compressed learning to a spatial sampling context by applying a binary mask \( \mathbf{B} \in \{0,1\}^{h \times w} \) directly to the image \( \mathbf{I} \in \mathbb{R}^{h \times w} \), producing compressed data through element-wise (Hadamard) multiplication:
\[
\mathbf{I}_{\text{compressed}} = \mathbf{B} \odot \mathbf{I}.
\]
Using this sampling method, our paper presents a CL framework for detecting deficiency values using a small subset of sampling points. This approach aims to overcome the long recording times caused by delays in the mechanical movement of the sensor in ARS.

\begin{figure*}[ht!]
\centerline{\includegraphics[scale=.65]{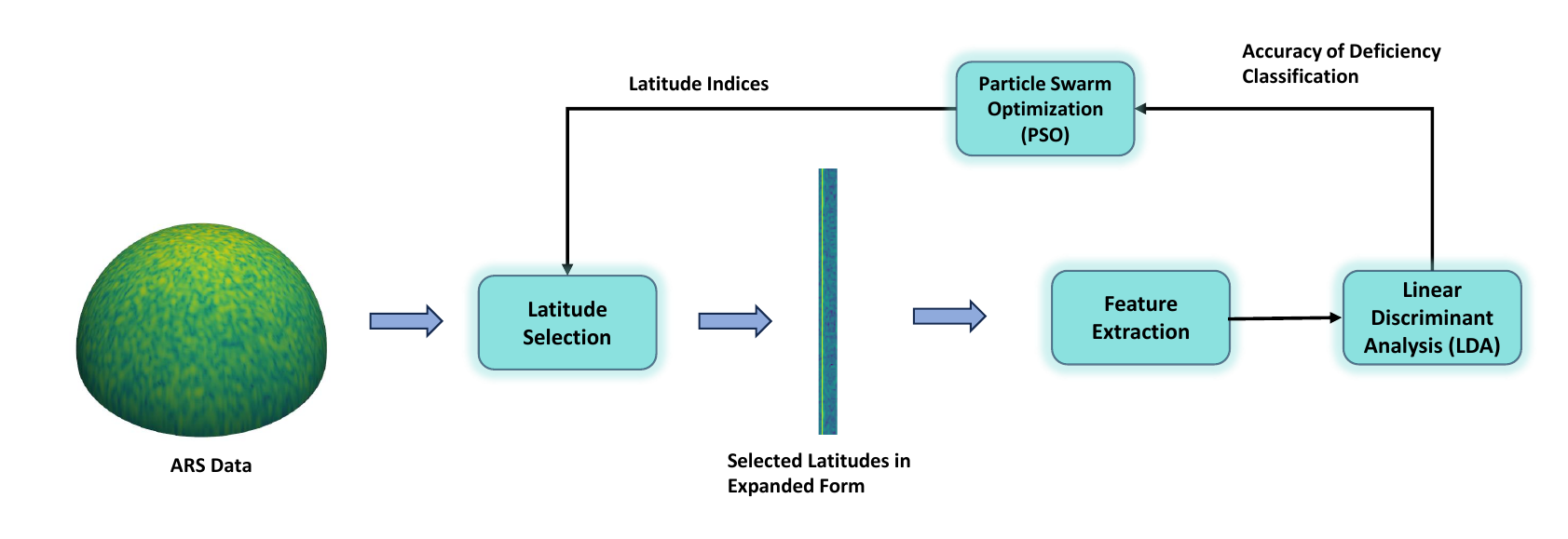}}
\caption{ Workflow of the proposed CL framework for latitudinal arc selection for the maximization of deficiency detection accuracy}
\label{fig1}
\end{figure*}

Fig.~\ref{fig1} shows the workflow of the proposed  CL framework.
In this framework, a particle swarm optimization (PSO) algorithm is employed alongside a linear discriminant analysis (LDA) classifier to search through the scattered light samples and identify the optimal combination that yields the highest accuracy for detecting deficiencies in ZnO nanosurface.


The main contributions of this work are as follows:
\begin{itemize}
\item A comprehensive CL framework is proposed for the analysis of ARS data, in which handcrafted features are extracted, important features are selected using recursive feature elimination (RFE), and classification is performed using linear discriminant analysis (LDA).
\item A latitude-based sampling scheme is introduced to reduce ARS acquisition time significantly by focusing on structurally informative sampling arcs while maintaining robustness to rotational variations.
\item The sampling optimization is defined as a constrained problem and solved using PSO, making it possible to identify the most informative sensing areas.
\item An intra-latitudinal sampling approach is further introduced, which enables a further reduction in sampling rate and enhances the system’s flexibility in detecting nanostructure deficiencies.
\end{itemize}

The remainder of this paper is organized as follows. Section II describes the proposed CL framework in detail. Section III presents the experimental evaluation of the framework under various conditions and reports the results. Section IV offers a deeper analysis and discussion of the proposed methodology. Finally, Section V summarizes the conclusions of this study.

\section{Methodology}
In the present section, we lay out the foundation of the proposed methodology by first introducing the dataset used in our work and the methods to provide a good representation of the dataset through handcrafted features and feature selection.
Then we continue by providing a more detailed description of the different parts of the proposed CL framework, including classification, optimization, and the sampling scheme. Additionally, we discuss how these components effectively integrate to form a functional framework for sampling and deficiency detection using ARS data.
Lastly, we extend the proposed sampling scheme to improve its flexibility and the performance of the framework in the detection of the deficiency level of the ZnO nanosurface using ARS images.

\subsection{ARS Dataset}
For our study, ARS data in different deficiency levels of ZnO nanosurfaces are needed. This will help us to evaluate the proposed method's accuracy in deficiency detection. Since, introducing a specific type of deficiency to a nanosurface is challenging \cite{robertson2012spatial}, simulated data will be used in our study.

For the simulation of the dataset, an open-source implementation of the DDA algorithm,called Amsterdam discrete dipole approximation (ADDA) \cite{yurkin2011discrete} is used. Scatterometry images of ZnO nanosurface with vacancy deficiencies of 0\%, 10\%, 20\%, 40\%, and 60\% are simulated.
ZnO nanosurface consists of vertically aligned nanorods that are randomly distributed but densely packed.  When a certain percentage of nanorods are absent, the surface morphology becomes irregular, and the overall density of the nanorods decreases. Vacancy deficiency percentages refer to the proportion of nanorods missing in the ZnO structure.
These changes in structural properties of ZnO will have effects on its interaction with light \cite{guo2024modulation} and subsequently on the resulting scatterometry data.


In our simulation configuration, a monochromatic laser beam with a wavelength of 405~nm is directed perpendicular to the nanosurface. The laser beam is Gaussian with a waist radius of 10~$\mu$m (defined as the $1/e^2$ radius). A light detector at a distance of 60~cm from the nanosurface scans the scattered light over the surrounding sphere \cite{rahman2024scatterometric}.
ZnO nanosurface is considered to be constructed on a transparent substrate, which makes it possible to have scattered light in the upper and lower hemispheres; in the upper hemisphere, we have reflected light, and in the lower hemisphere, transmitted light.

We only used reflected scattered light with a light detector, scanning the upper hemisphere with an angular resolution of \(1^\circ\). The angles are \(\alpha\), the vertical angle measured from the equator $(0^\circ)$ to the pole $(90^\circ)$ and back to the equator $(180^\circ)$, and \(\beta\), the horizontal angle measured around in a semicircle. \(\beta\) could range from $0^\circ$ to \(180^\circ\) or $0^\circ$ to \(-180^\circ\)(see Fig. \ref{fig_angles}):

\[
\alpha \in [0^\circ, 180^\circ], \quad \beta \in [-180^\circ, 180^\circ]
\]

\begin{figure}[hbtp]
\centerline{\includegraphics[width=0.6\columnwidth]{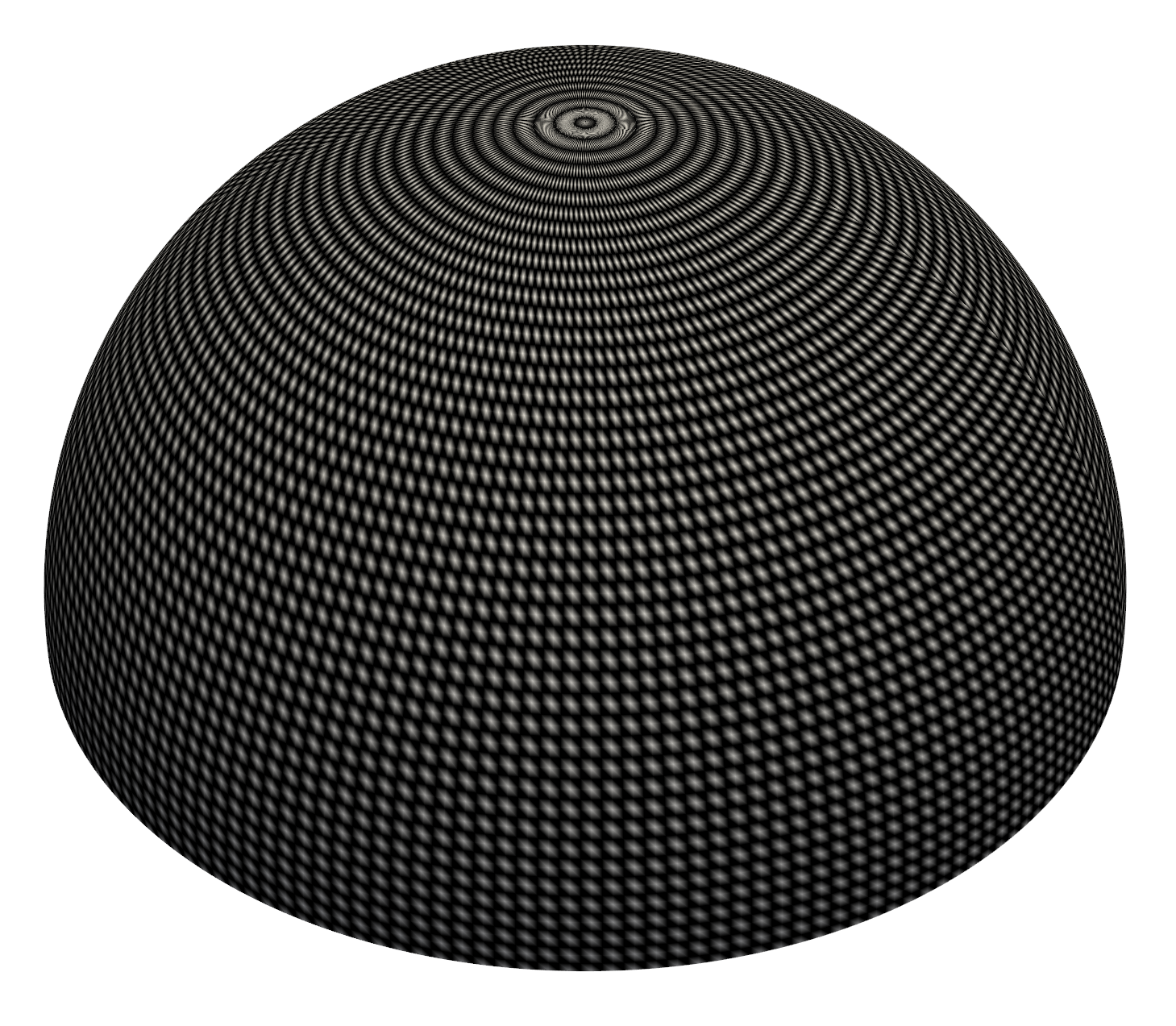}}
\caption{Scanning positions of the light-detecting sensor for a single ARS image acquisition}
\label{fig_sam}
\end{figure}

To have a better understanding of the sampling points Fig.~\ref{fig_sam} is provided, which shows the possible positioning of the sensor on the sampling sphere.

\begin{figure}[hbt!]
\centerline{\includegraphics[width=\columnwidth]{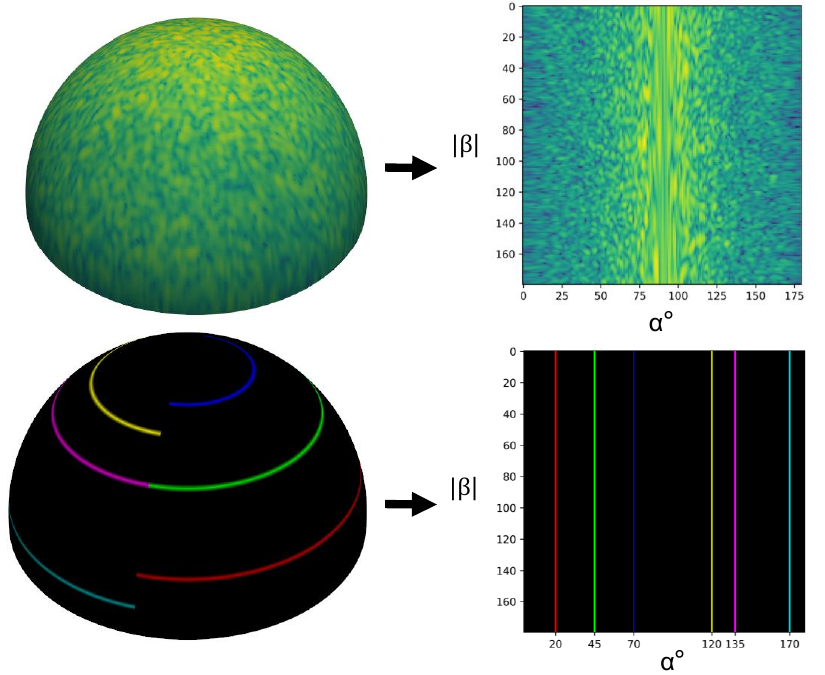}}
\caption{An illustration of ARS data in spherical form and its transformation to the expanded 2D image. In the second row, there is an illustration of the placement of latitudinal arcs in the corresponding columns in the 2D image.}
\label{fig2}
\end{figure}

A set of 800 scatterometry images, each with a size of $180 \times 180$, was created to represent five distinct levels of deficiency. 
Within this dataset, there are 160 images corresponding to each deficiency level. Fig.~\ref{fig2}, the first row illustrates an example of an ARS image. For demonstration purposes, the spherical image is converted into a 2D image.
In this expanded view, the half-latitudes of the sphere are arranged along the columns of the 2D image, which can be seen in Fig.~\ref{fig2}, bottom row. The details regarding these half-latitudes will be discussed in Section \ref{subsec:Definition of Latitude-based Sampling Scheme}. 

\subsection{Feature Extraction, Feature Selection and Classification}


Here, as the first part of the CL framework, we introduce the classification model and handcrafted features for the discrimination between ARS images with different deficiency levels.

Through feature extraction, high-dimensional data can be transformed into meaningful representations. This process reduces complexity while preserving essential information. 
The extracted features reflect patterns, trends, and relationships within the data which makes interpretation and analysis easier.
In the context of ARS data, selecting and extracting key features help in distinguishing between different deficiency levels of nanosurfaces.

Due to the novelty of ARS imaging, a suitable feature space that could thoroughly represent ARS data has not yet been explored in the literature. 
Therefore, we first conduct a comprehensive feature extraction to have a large set of features representing ARS images from different perspectives. 
Then, we utilize a feature selection method to select specific features that are contributing to the detection of deficiencies. The extracted features at the first stage can be categorized into basic statistical features, higher-order statistical features, histogram-based features, and texture-based features.
Statistical features can help detect overall ARS image intensity distribution.
Histogram-based features are useful in analyzing ARS image intensities and contrast variations without considering spatial relationships.
Texture-based features provide information on speckle patterns in ARS images.
The extracted features build a feature space that presents images in a lower dimension. 

\begin{figure}[ht!]
\centerline{\includegraphics[scale=.9]{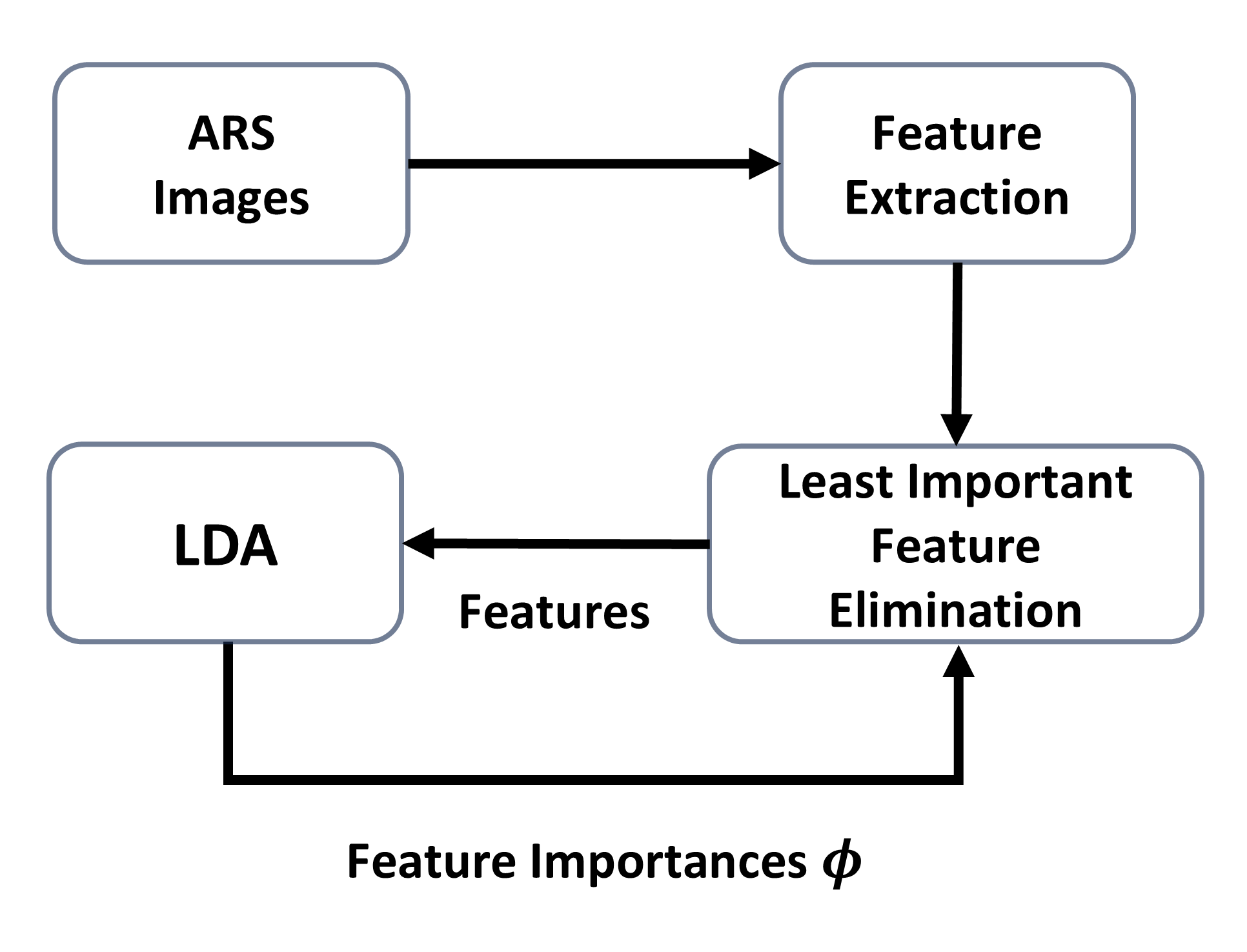}}
\caption{Workflow of recursive feature elimination (RFE) with linear discriminant analysis (LDA)-based feature importance}
\label{fig33}
\end{figure}

In order to further reduce this feature space, eliminate redundancy, retain the most significant features, and speed up feature extraction, we use the recursive feature elimination (RFE) method. In RFE, features are recursively removed according to their importance until the optimal subset is found. 
RFE follows the objective of improving the performance of a machine learning model by eliminating the least important feature one by one. To achieve this, the machine learning model must be capable of assigning feature importance scores. The process of feature selection in RFE is also shown in Fig.~\ref{fig33}. 

In our study, linear discriminant analysis (LDA) is used as the machine learning model.
For the detection of deficiencies of nano surfaces from ARS data with different deficiency levels, we employ the LDA as a classifier. In addition, LDA can also provide a feature importance score for the feature vector in the input space, which then will be used as an input for RFE to reduce the number of features. 
LDA classifier is a supervised machine learning algorithm. It works by modeling the distribution of each class, assuming a Gaussian distribution with the same covariance matrix but different means. LDA projects the data onto a lower-dimensional space by finding a linear combination of features that best separates the classes. It maximizes the distance between the means of different classes while minimizing the variance within each class. 
LDA achieves that by optimizing Fisher’s criterion, which maximizes the ratio of between-class variance to within-class variance. This projection creates well-separated classes while maintaining tight clustering within each class. As a result, the algorithm produces linear decision boundaries that effectively distinguish between the classes.

The following presents the mathematical formulation of LDA, detailing Fisher’s criterion, the underlying mechanism for class separation, and how feature importance is derived from the projection matrix.

Given a dataset of \( N \) samples (feature vectors) \( \mathbf{f}_n \in \mathbb{R}^d \), each belonging to one of \( k \) classes, the matrices that characterize class separability in LDA are defined as follows:


The between-class scatter matrix is defined as

\begin{equation}
S_B = \sum_{c=1}^{k} N_c (\boldsymbol{\mu}_c - \boldsymbol{\mu})(\boldsymbol{\mu}_c - \boldsymbol{\mu})^T
\end{equation}

\noindent where \( \boldsymbol{\mu}_c \) is the mean vector of class \( c \), \( \boldsymbol{\mu} \) is the overall mean vector, and \( N_c \) is the number of samples in class \( c \).

The within-class scatter matrix is defined as

\begin{equation}
S_W = \sum_{c=1}^{k} \sum_{\mathbf{f} \in C_c} (\mathbf{f} - \boldsymbol{\mu}_c)(\mathbf{f} - \boldsymbol{\mu}_c)^T
\end{equation}

\noindent where $ C_c $ represents the set of samples in class \( c \).

The optimal projection matrix  \( W \in \mathbb{R}^{d \times (k - 1)} \) that maximizes class separability is obtained by solving the generalized eigenvalue problem:

\begin{equation}
S_W^{-1} S_B W = \Lambda W
\end{equation}

\noindent where \( \Lambda \) is the diagonal matrix of eigenvalues, and the eigenvectors in \( W \) define the transformation that maximizes class separability.

Feature importance in LDA can be derived from the absolute values of the elements in the projection matrix \( W \). Specifically, the importance of the \( j \)-th feature can be estimated using:

\begin{equation}
\phi_j = \sum_{c=1}^{k-1} | W_{jc} |
\end{equation}

\noindent where \( W_{jc} \) represents the weight assigned to feature \( j \) in the discriminant function. A higher magnitude indicates that the feature contributes more to class separation.

Using RFE and feature importances, a comprehensive feature set of 30 features is reduced to 11 without any loss in the LDA's classification performance.

The selected features are shown in Table~\ref{tab00}. In this table, details of the calculation of each feature for a given input signal or image $\mathbf{I}_n$ are given.
Each image \( \mathbf{I}_n \) contains \( M \) pixels, denoted as \( \mathbf{I}_{n,i} \), where \( i = 1, \ldots, M \).
Equation~\ref{eq:f} shows that all of these features form a feature vector $\mathbf{f}_n = f(\mathbf{I}_n)$  that provides a specific representation of the scatterometric image $\mathbf{I}_n$ and helps to distinguish scatterometric patterns that are present in the ARS data.

\begin{align}
f(\mathbf{I}_n) = &\, [ \left.  \mu(\mathbf{I}_n), \, \sigma (\mathbf{I}_n),  \sigma^2(\mathbf{I}_n),   \min(\mathbf{I}_n), \, \tilde{\mathbf{I}}_n, \, P_{25}(\mathbf{I}_n), \right. \nonumber \\
          & \left.  \, P_{10}(\mathbf{I}_n), \, \text{MAD}(\mathbf{I}_n),  \text{RMS}(\mathbf{I}_n), \, \text{E}(\mathbf{I}_n), \right. \nonumber \\
          & \left. \, \text{R}(\mathbf{I}_n)  ] \right.  \,  
\label{eq:f}
\end{align}


    

\begin{table}[h]
    \caption{The selected features by RFE}
    \centering
    \renewcommand{\arraystretch}{1.9}
    \begin{tabular}{p{3.8cm} p{3.5cm}}
        \textbf{Feature} & \textbf{Calculation} \\ 
        \hline
        Mean ($\mu$) & $\frac{1}{M} \sum_{i=1}^{M} \mathbf{I}_{n,i}$ \\ 
        \hline
        Standard Deviation ($\sigma$) & $\sqrt{\frac{1}{M} \sum_{i=1}^{M} (\mathbf{I}_{n,i} - \mu)^2}$ \\ 
        \hline
        Variance ($\sigma^2$) & $\sigma^2$\\ 
        \hline
        Minimum Value ($\min$) & $\min(\mathbf{I}_{n,1}, \mathbf{I}_{n,2}, ..., \mathbf{I}_{n,M})$ \\ 
        \hline
        Median Value ($\tilde{\mathbf{I}}_n$) & Middle value of sorted data \\ 
        \hline
        25th Percentile ($p_{25}$) & Value at $25 \%$ of sorted data \\ 
        \hline
        10th Percentile ($p_{10}$) & Value at $10 \%$ of sorted data \\ 
        \hline
        Mean Absolute Deviation (MAD) & $\frac{1}{M} \sum_{i=1}^{M} |\mathbf{I}_{n,i} - \mu|$ \\ 
        \hline
        Root Mean Square (RMS) & $\sqrt{\frac{1}{M} \sum_{i=1}^{M} \mathbf{I}_{n,i}^2}$ \\ 
        \hline
        Energy (E) & $\sum_{i=1}^{M} \mathbf{I}_{n,i}^2$ \\ 
        \hline
        Range (R) & $\max(\mathbf{I}_{n}) - \min(\mathbf{I}_{n})$ \\ 
    \end{tabular}
    
    \label{tab00}
\end{table}




\subsection{Optimization of ARS Sampling}

To improve the recording time of the ARS, a subset of sampling points that are most contributing to the detection of deficiencies will be scanned instead of the entire sphere.
The ARS device has the capability to sample every single point in the scanning sphere independently. Therefore, it is possible to define a search space that includes combinations of different points in the scanning sphere or different pixels in the ARS data. 

However, there are two major drawbacks to using a single point-based search space. Firstly, it needs a significant amount of time to find the best combinations of the pixel positions leading to the most accurate deficiency detection result.
Secondly, due to the spherical nature of the data, the point-based sampling will be sensitive to the rotation around the axis of the incident light (shown as the z-axis in Fig. \ref{fig_angles}). This sensitivity is problematic because the optimizer can converge to the sampling patterns that are not related to the structural properties of the nanosurface but the placement of it, and any changes in the placement or the rotation of the nanosurface can cause the previously found sensing pattern to be invalid.

To address the challenges mentioned, we propose a latitude-based sampling scheme to effectively sense ARS data. 
In this approach, the search space is limited to specific latitudinal arcs. This significantly reduces the search space of the optimizer that searches for a specific set of arcs. By limiting search space, the optimizer can focus its efforts on more targeted regions and can effectively search for the optimum sampling areas. 
Furthermore, the latitude-based sampling scheme is inherently robust to rotations around the z-axis, which results in an optimization that is not sensitive to such rotations.

\subsubsection{Definition of the Latitude-based Sampling Scheme}
\label{subsec:Definition of Latitude-based Sampling Scheme}

In this scheme, a set of latitudinal arcs is considered to be sampled for analyzing the scattering properties of the surface under study.

\begin{figure}[hbt!]
\centerline{\includegraphics[width=0.7\columnwidth]{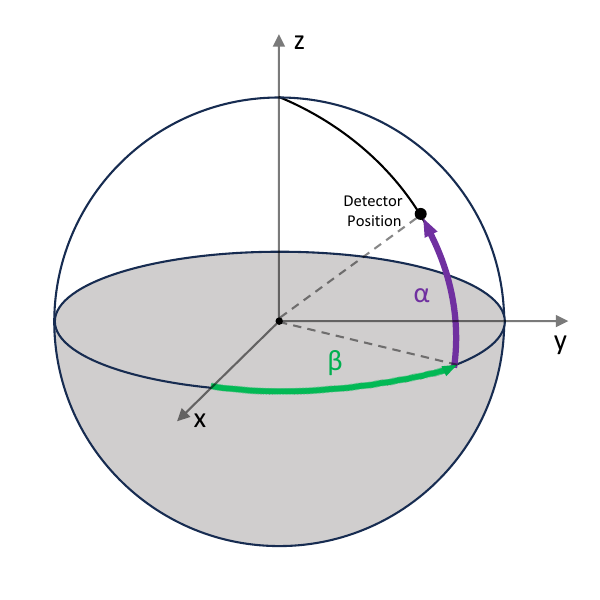}}
\caption{Vertical angle $(\alpha)$ and horizontal angle $(\beta)$ in ARS  sphere}
\label{fig_angles}
\end{figure}

In our study, each latitudinal arc, or arc in short, is defined in such a way that it spans half of a sphere's latitude at a specific angle. 
In other words, a single latitude is split into two arcs, the first arc spanning horizontal angle $ \beta $ from 0 to 180 degrees and vertical angle $\alpha = \alpha_1^\circ$, $\alpha_1$ being a specific angle between $0^\circ$ and $90^\circ$. Horizontal and vertical angles are shown in Fig. \ref{fig_angles} for clarity.
The second arc spanning horizontal angle $ \beta $ from $0^\circ$ to $-180^\circ$ degrees but with vertical angle $\alpha_2$, where $\alpha_2 = 180^\circ - \alpha_1^\circ$.

For instance, a few of these latitudinal arcs are illustrated in the bottom row of Fig.~\ref{fig2}. In this figure, six latitudinal arcs, $\alpha = [20^\circ, 45^\circ, 70^\circ, 135^\circ, 120^\circ, 170^\circ]$ with distinct colors are shown in both spherical and expanded form. 
The reason for considering half-latitudes or latitudinal arcs instead of full latitudes is that it provides more flexibility in sampling to search through different latitudes with a lower number of sampling points.

The motivation behind this sampling scheme arises from the observation that different latitudinal arcs of the sphere, or alternatively, different columns of the 2D image, show distinct scattering patterns. For instance, high-intensity light is present at the pole of the sphere in contrast to the equator area, which has lower intensities as illustrated in Fig.~\ref{fig2} first row.
By using latitudinal arc selection, we ensure that the scanned data provide the required information about the scattering patterns.
The search space is defined as the range of latitudinal arcs covering the possible angles of reflection in the upper hemisphere of the ARS data. This method enables a thorough search of the angular domain.

Let $ \mathbf{\Lambda} = \{\alpha_1, \alpha_2, \dots, \alpha_k\} $ be the subset of arcs from the full set of available arcs \( [0^\circ, 180^\circ] \).

For each latitudinal arc subset $ \mathbf{\Lambda} $, the feature extraction function $ f(\mathbf{\Lambda}) $ computes the statistical features.

The goal is to find the subset $ \mathbf{\Lambda}^* $ that maximizes the classification accuracy $ \text{acc}(f(\mathbf{\Lambda})) $ of the LDA classifier. This can be expressed as:

\begin{equation}
\label{eq1}
\mathbf{\Lambda}^* = \operatorname*{arg\,max}_{\mathbf{\Lambda} \subset [0^\circ, 180^\circ]} \text{acc}(f(\Lambda)) \quad \text{s.t.} \quad |\mathbf{\Lambda}| \leq k
\end{equation}

\noindent where $ k $ is a predefined maximum number of latitudinal arcs to sample. We impose a constraint on the size of the arc set $ (|\mathbf{\Lambda}| )$ to control the sampling rate.

In this formulation, finding an exact solution to this problem could be considered NP-hard due to the exponential growth in the number of possible combinations of scanning points.
Exhaustive search methods can be used to find the optimum solution. However, due to the high complexity of the problem, these approaches can take a significant amount of time to find the solution.
A more efficient method with faster convergence includes the greedy search method that finds a suboptimal solution. However, this method has the drawback of converging to local optima \cite{wu2018beyond}.
Considering the mentioned challenges, we will proceed with a metaheuristic optimization algorithm called particle swarm optimization (PSO).

\subsubsection{Particle Swarm Optimization}

We aim to use PSO \cite{kennedy1995particle} as the optimizer for the optimization problem \eqref{eq1}. PSO is known for its robustness in diverse applications and fast convergence \cite{wang2018particle}. Inspired by the social behavior of swarms, PSO creates a group of particles that explore the search space. These particles represent potential solutions to the problem. At first, particles are randomly spread out throughout the search space. At each iteration, each particle changes its position based on its own experience and the experiences of nearby particles to find the best position.
This means that particles are influenced both by the best position they have found so far, known as personal best, and by the best position found by any particle in the swarm, known as global best.

In PSO, the velocity of each particle determines its movement in the search space. The velocity vector  $\mathbf{v}_{k}(i)$ represents the step size and direction of particle $k$ at iteration $i$.
The update of each particle's velocity is controlled by three parameters: inertia $w$, cognitive $c_1$, and social $c_2$. The inertia parameter controls the effect of the particle's current velocity in the next velocity update, the cognitive parameter determines the impact of the personal best position in the velocity update, and the social parameter determines the effect of the global best position in the velocity update.

The velocity update equation in PSO can be expressed as:

\begin{equation}
\begin{split}
\mathbf{v}_{k}(i+1) = w \cdot \mathbf{v}_{k}(i) 
+ c_1 \cdot r_1 \cdot (\mathbf{p}_{k}^{pbest} - \mathbf{p}_{k}(i)) \\
+ c_2 \cdot r_2 \cdot (\mathbf{p}^{gbest} - \mathbf{p}_{k}(i))
\end{split}
\label{eq:v_update}
\end{equation}


\noindent where $r_1$ and $r_2$ are random numbers between $0$ and $1$, $\mathbf{p}_{k}^{\text{pbest}}$ is the personal best position of particle $k$, $\mathbf{p}^{\text{gbest}}$ is the global best position found by the swarm, and $\mathbf{p}_{k}(i)$ is the current position of particle $k$.
Random factors $r_1$ and $r_2$ are included in the PSO to balance exploration and exploitation inside the search space.


In the context of latitude-based sampling, each particle in the PSO algorithm represents a set of sampling arcs. The degrees or positions of these arcs within the ARS sphere serve as optimization parameters that the particles try to adjust.
From a geometric perspective, the number of dimensions in the search space corresponds to the number of arcs being sampled, and the position of each particle within this space defines the positions of the arcs. For instance, after an iteration when a particle has positioned itself in the search space, the location of a specific arc on the ARS sphere can be determined by projecting a particle onto the corresponding axis in the search space.

At each iteration of the optimization, the feature extraction is carried out on the sampled arcs by the PSO algorithm. Then, the extracted features are used to train and test the LDA classifier, which evaluates their effectiveness for classification. 
The classification accuracy obtained from this evaluation is subsequently fed back into the PSO algorithm, as shown in Fig.\ref{fig1}. This feedback mechanism enables the algorithm to adjust the latitudinal arc positions in the next iteration and improve the sampling process to focus on the most promising regions.
This iterative process continues until the algorithm converges to the set of arcs with the highest classification accuracy.
A detailed step-by-step description of the CL framework is provided in the algorithm \ref{alg}.


\begin{algorithm}
\caption{Compressed Learning Framework for ARS Image Classification}\label{alg}
\begin{algorithmic}
    \STATE \textbf{Input:} A set of ARS images $\mathbf{I}^{180 \times 180}$, \\
            \hspace{2.7em}       Number of latitudinal arcs to sample $L$\\
            \hspace{2.7em}       Swarm size or number of particles: $K$
    \STATE \textbf{Output:} Optimized sampling arc indices ($\mathbf{p}_{\text{gbest}}$), \\
            \hspace{3.5em} Optimized classification accuracy $\mathbf{acc}_{\text{gbest}}$
    \STATE Initialize  particle positions $\mathbf{P}^{L \times K}$ randomly
    \STATE Initialize empty set of features $\mathbf{F}$
    \FOR{ $i = 1: i \gets i+1$ until stopping criteria met}
        \FOR{each particle $k$ with position $\mathbf{p}_k$ in the swarm}
            \FOR{ all ARS images in the set}
                \STATE $\mathbf{\Lambda}$ = $\mathbf{I}[\mathbf{:,p}_{k}]$ 
                \STATE Add $f(\mathbf{\Lambda})$ (Eq.~\ref{eq:f}) to feature set $\mathbf{F}$

            \ENDFOR
            \STATE $\mathbf{F}_{\text{train}}, \mathbf{F}_{\text{test}} \leftarrow   \text{Random Split}(\mathbf{F})$            
            \STATE $\mathbf{acc}_{\text{train}} \leftarrow$  \text{LDA}($\mathbf{F}_{\text{train}}$)
             \IF{$\mathbf{acc}_{\text{train}} > \mathbf{acc}_{\text{pbest}, k}$}
            \STATE  $\mathbf{p}_{\text{pbest},k} \leftarrow \mathbf{p}_{k}$
            \STATE $\mathbf{acc}_{\text{pbest}, k} \leftarrow \mathbf{acc}_{\text{train}} $
             \ENDIF
              \IF{$ \mathbf{acc}_{\text{train}}   > \mathbf{acc}_{\text{gbest}}$}
             \STATE  $\mathbf{p}_{\text{gbest}} \leftarrow \mathbf{p}_{k}$
             \STATE $\mathbf{acc}_{\text{gbest}} \leftarrow \mathbf{acc}_{\text{train}} $
             \ENDIF
             \STATE Update velocity $\mathbf{v}_{k}$ using Eq.~\ref{eq:v_update}

             \STATE Update particle position $\mathbf{p}_{k} = \mathbf{p}_{k} + \mathbf{v}_{k}$ \\
             \STATE Discretize $\mathbf{p}_{k}$ to valid arc indices

        \ENDFOR
    \ENDFOR

\end{algorithmic}
\end{algorithm}

\subsection{Extension to Intra-latitudinal Sampling}
To further extend the capability of the methodology to reach lower sampling rates and give more flexibility to the sampling process, we extend the proposed method to have intra-latitudinal sampling on the ARS sampling sphere. In this scheme, after having the sampled latitudinal arcs from the original latitude-based sampling, we run the algorithm for the second time with a few modifications. In the second run, the optimization algorithm takes the selected arcs of the ARS images as the input. Then, it reduces the number of sampling points within arcs according to a chosen sampling rate. 

In this process, the particles in the PSO algorithm originally responsible for optimizing latitudinal sampling are now tasked with identifying the optimal combination of sampling points within the previously optimized latitudinal arcs. This hierarchical optimization approach ensures that the final set of sampling points is not only sparse but also strategically distributed. 

The intra-latitudinal sampling scheme ensures the minimization of redundancy. This is because uniform or high-density sampling often leads to redundant data points, where neighboring samples contain similar or repetitive information. By optimizing both latitude-based and intra-latitudinal sampling, our approach effectively eliminates unnecessary points, thereby  reducing sampling time and the computational complexity of downstream data processing while preserving essential information.

\section{Experimental results}
In this section, we evaluate the proposed framework through a series of experiments designed to assess its performance under various conditions.
We aim to validate the suitability of our framework in finding areas within ARS data patterns that are most beneficial for the detection of deficiency levels.

For all experiments, PSO is configured with the following parameters: an inertia weight of \( w = 0.5 \), a cognitive coefficient of \( c_1 = 1 \), and a social coefficient of \( c_2 = 3 \). These values were selected to balance exploration and exploitation in the search space. The algorithm is run for 80 iterations  to allow the model to converge to the optimal solution.
Additionally, to have reproducible results in all of our experiments, we have repeated the experiment ten times, and the reported results are the average over all repetitions.

\subsection{Deficiency Detection Using Full ARS Data}
\label{3A}
We first study the suitability of the feature extraction and classification methods with fully sampled data. This also provides us with benchmark results to further compare with the compressively sensed samples.
To better model the data in real-world scenarios, different levels of noise are introduced to the data. The introduced noises include Gaussian noise with a signal-to-noise ratio (SNR) of $20$~dB and $30$~dB to model the mechanical vibrations in the device and detector sensitivity, and salt-and-pepper noise with $10\%$ corruption to model sensor transmission error. Additionally, three classifiers are employed for the detection of deficiencies across five different classes.

The results, presented in Table~\ref{tab0}, clearly demonstrate that the extracted features could provide a distinctive representation of the ARS data to distinguish different levels of deficiencies in nanosurfaces. As for the noiseless case, the linear discriminant analysis (LDA) classifier achieved almost perfect classification results with an accuracy of $97.58\%$. Further, when noise is introduced, the LDA continues to perform robustly.
Additionally, two other widely used classifiers, Support Vector Machine (SVM), which works by finding the optimal hyperplane to separate classes, and K-Nearest Neighbors (KNN), which classifies data points based on the majority class of their nearest neighbors, are used for comparison.
However, the LDA classifier outperforms the SVM and KNN, as they both fail to detect surface deficiencies with salt-and-pepper noise. This makes LDA the preferred choice in the CL framework.

\begin{table}[htbp]
\scriptsize
\caption{Test accuracy of different classifiers for detecting nanosurface deficiency using full ARS data with different levels of Gaussian and salt-and-pepper noise}
\begin{center}
\renewcommand*{\arraystretch}{2}
\begin{tabular}{c c c c }
\hline 
  &\multicolumn{3}{c}{Classification Model} \\
\cline{2-4} 
\textbf{Noise} & LDA& SVM& KNN\\
\hline
Noiseless &  97.58\%  & 74.38& 77.50\% \\
\hline
Gaussian with 20 dB SNR &  94.62 \% & 74.38\% & 75.62\%   \\
\hline
Gaussian with 30 dB SNR &  95.50 \% & 75.18\% & 76.88\%   \\
\hline
Salt-and-Pepper (10\%) & 96.50\% & 15\% & 41.88\%  \\
\hline
Gaussian (20 dB) and Salt-and-Pepper (10\%) & 85.0\% & 15\% & 44.0\%  \\
\hline
\end{tabular}
\label{tab0}
\end{center}
\end{table}


\subsection{Latitude-Based Sampling Scheme}

\begin{figure}[hbt!]
\centerline{\includegraphics[width=1.1\linewidth]{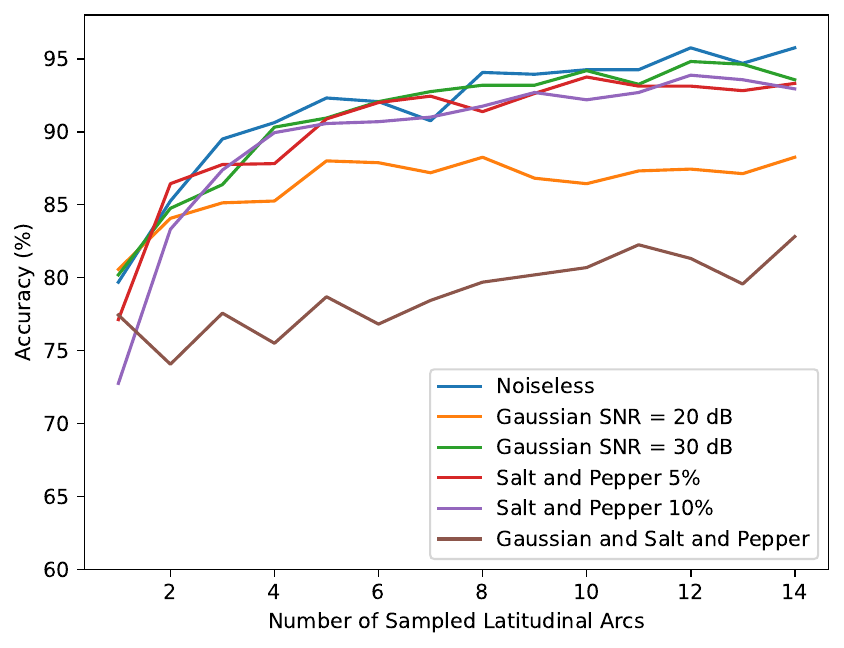}}
\caption{ Achieved deficiency detection accuracy using the CL framework with different numbers of sampled latitudinal arcs and different levels of Gaussian and salt-and-pepper noise in ARS data}
\label{fig3}
\end{figure}

In this experiment, we study the effect of selecting different numbers of latitudinal arcs on the deficiency detection results.
Fig.~\ref{fig3} shows the deficiency detection accuracies with different numbers of sampling arcs.
After specifying the number of arcs for sampling, the CL framework iterates and searches for a combination of arcs that maximizes classification accuracy.
The framework is evaluated with different numbers of sampling arcs ranging from $1$ to $14$ out of $180$ latitudinal arcs, corresponding to sampling rates ranging from $0.5\%$ to $7.7\%$.

For instance, in a noiseless scenario with only $12$ arcs, the classification accuracy reaches $95\%$, whereas the full image classification accuracy stands at $97.5\%$.
This means that reducing the sampling points by $95\%$ only decreases the classification accuracy by $2.5\%$.
It can also be seen that the noiseless curve is closely followed by Gaussian noise with a $30$~dB SNR and $10\%$ salt-and-pepper noise, which indicates the robustness of the CL framework in noisy conditions.
The framework also shows great potential by achieving $84.5\%$ of deficiency detection using just one latitudinal arc, or $180$ data points out of a total of $32400$ points.

\subsection{Sensing Areas}

After observing the framework's success in identifying the most useful latitudinal arcs for deficiency detection, we are now interested in determining which areas of the recording sphere are frequently utilized by the CL framework. In other words, we aim to find the regions that contain critical information for distinguishing ARS images of nanosurfaces with various deficiency levels.

\begin{figure}[hbt!]
\centerline{\includegraphics[width=\columnwidth]{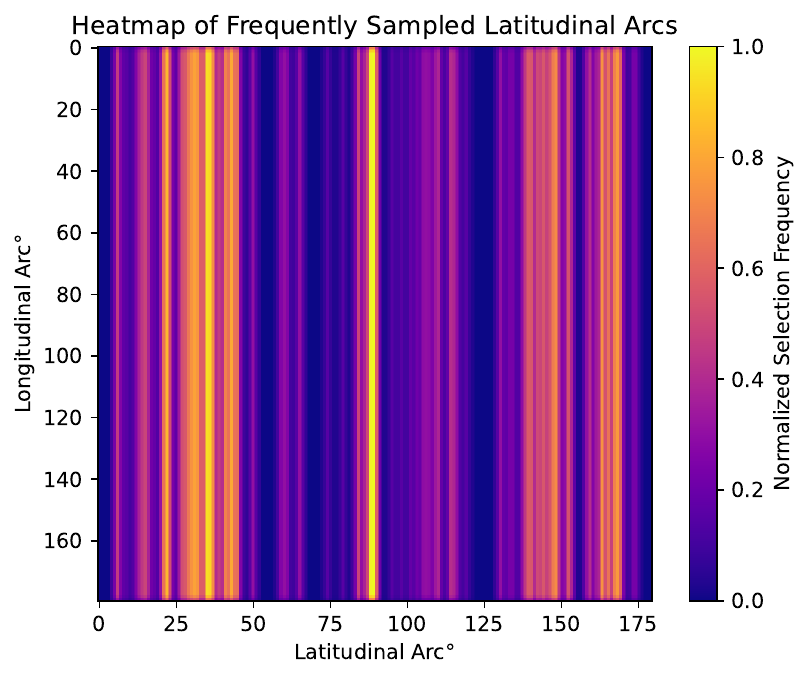}}
\caption{Heatmap of Frequently Sampled Latitudinal Arcs}
\label{fig4}
\end{figure}

Fig.~\ref{fig4} displays a heatmap generated from the sampled latitudinal arcs over 20 different runs, representing the most frequently sampled latitudinal  arcs through color intensity.
The heatmap is plotted in an expanded form of the sphere as a 2D image.
In this heatmap, the center line corresponds to the region around the poles of the sphere, while the sides of the image correspond to the regions around the equator, as illustrated in Fig.~\ref{fig2}.
The lighter regions in the heat map are sampled the most, and the darker regions are the less frequently sampled regions.

Two specific regions are selected the most during the CL process.
First, the region near the poles, which was expected because the light source is placed in this area directly over the nanosurface. Therefore, most of the incident light reflects back at the same angle as the incident angle. Consequently, the area around the pole has higher light intensities, as can be seen in a sample image in Fig.~\ref{fig2}.
Second, the regions on two sides of the heat map or around the equator, which include latitudinal arcs in the range of $0^\circ$ to $50^\circ$ and $130^\circ$ to $180^\circ$. We can see from the sample images that these regions include light intensities with higher frequency and sharp intensity changes.
Because this region has a larger area, the reflected light is not superimposed, and the scattering patterns are visible. This might be a possible reason for the equatorial regions to be of interest to the CL framework.

\subsection{Intra-latitudinal Sampling Scheme}

In the designed experiment for intra-latitudinal sampling, we test the performance of the proposed method for different numbers of latitudinal arcs and also different sampling rates. In this experiment, first the latitude-based sampling is carried out with a specific number of arcs, and after having the optimized arcs, the intra-latitudinal sampling takes over and reduces the number of sampling points with a specific rate.

The accuracy values for the classification of 5 different deficiency levels in ARS images with a limited number of sampling points are reported in Table~\ref{tab3}. From left to right, the number of sampled arcs increases from $2$ to $15$, while moving down the rows corresponds to an increase in the sampling rate from $5\%$ to $90\%$. The values inside each cell of the table are the classification accuracy values in bold in the upper part of the cell, and the number in the lower part of the cell represents the number of sampling points out of 32400 points in the original ARS sphere. For instance, the last value in the first row shows 81.90\% classification accuracy using $5\%$ of $15$ latitudinal arcs of the ARS data in which only $135$ data points are used.

Examining Table~\ref{tab3} from a different viewpoint shows the robustness of the combined latitudinal and intra-latitudinal sampling methods for a wide range of configurations. Even at the minimal settings with fewer than 100 sampling points, the accuracy values can reach up to 83.3\%. This indicates the method's ability to extract meaningful information from extremely sparse data.
As the number of sampling points increases—for example, when using $7$ arcs at a $50\%$ sampling rate, which results in $630$ points—the method maintains strong performance, achieving an accuracy of $89.40\%$. 
This adaptability suggests that the proposed method performs well in exploiting critical characteristics of the ARS images and makes it highly effective for applications where sampling flexibility and resource efficiency are important.



An important aspect to consider is the robustness of the proposed framework across different sampling rates and sampling schemes. This consistency mostly results from the selected features, because the chosen features are mostly statistical and value-based, meaning they rely on individual data point values rather than underlying structural patterns of the ARS data.

\begin{table*}[h!]
\centering
\renewcommand{\arraystretch}{2}
\setlength{\tabcolsep}{6pt}
\begin{tabular}{c*{14}{c}}
\multicolumn{15}{c}{\textbf{Number of Sampled Latitudinal Arcs}} \\
\hline
\textbf{Sampling Rate (\%)} & 2 & 3 & 4 & 5 & 6 & 7 & 8 & 9 & 10 & 11 & 12 & 13 & 14 & 15 \\ \hline
\textbf{5}  & \makecell{\textbf{54.38} \\ \scriptsize{18}}  & \makecell{\textbf{69.75} \\ \scriptsize{27}}  & \makecell{\textbf{75.50} \\ \scriptsize{36}}  & \makecell{\textbf{74.20} \\ \scriptsize{45}}  & \makecell{\textbf{76.20} \\ \scriptsize{54}}  & \makecell{\textbf{77.70} \\ \scriptsize{63}}  & \makecell{\textbf{78.25} \\ \scriptsize{72}}  & \makecell{\textbf{80.25} \\ \scriptsize{81}}  & \makecell{\textbf{83.30} \\ \scriptsize{90}}  & \makecell{\textbf{82.56} \\ \scriptsize{99}}  & \makecell{\textbf{82.00} \\ \scriptsize{108}} & \makecell{\textbf{80.94} \\ \scriptsize{117}} & \makecell{\textbf{82.75} \\ \scriptsize{126}} & \makecell{\textbf{81.90} \\ \scriptsize{135}} \\ \hline
\textbf{10} & \makecell{\textbf{63.30} \\ \scriptsize{36}}  & \makecell{\textbf{75.50} \\ \scriptsize{54}}  & \makecell{\textbf{79.80} \\ \scriptsize{72}}  & \makecell{\textbf{79.70} \\ \scriptsize{90}}  & \makecell{\textbf{80.06} \\ \scriptsize{108}} & \makecell{\textbf{80.75} \\ \scriptsize{126}} & \makecell{\textbf{80.90} \\ \scriptsize{144}} & \makecell{\textbf{83.75} \\ \scriptsize{162}} & \makecell{\textbf{83.75} \\ \scriptsize{180}} & \makecell{\textbf{84.60} \\ \scriptsize{198}} & \makecell{\textbf{83.40} \\ \scriptsize{216}} & \makecell{\textbf{84.44} \\ \scriptsize{234}} & \makecell{\textbf{86.06} \\ \scriptsize{252}} & \makecell{\textbf{86.30} \\ \scriptsize{270}} \\ \hline
\textbf{20} & \makecell{\textbf{75.56} \\ \scriptsize{72}}  & \makecell{\textbf{80.50} \\ \scriptsize{108}} & \makecell{\textbf{83.00} \\ \scriptsize{144}} & \makecell{\textbf{83.06} \\ \scriptsize{180}} & \makecell{\textbf{85.06} \\ \scriptsize{216}} & \makecell{\textbf{85.60} \\ \scriptsize{252}} & \makecell{\textbf{86.10} \\ \scriptsize{288}} & \makecell{\textbf{86.25} \\ \scriptsize{324}} & \makecell{\textbf{87.10} \\ \scriptsize{360}} & \makecell{\textbf{87.56} \\ \scriptsize{396}} & \makecell{\textbf{87.06} \\ \scriptsize{432}} & \makecell{\textbf{90.06} \\ \scriptsize{468}} & \makecell{\textbf{88.44} \\ \scriptsize{504}} & \makecell{\textbf{90.20} \\ \scriptsize{540}} \\ \hline
\textbf{30} & \makecell{\textbf{76.94} \\ \scriptsize{108}} & \makecell{\textbf{82.30} \\ \scriptsize{162}} & \makecell{\textbf{84.50} \\ \scriptsize{216}} & \makecell{\textbf{86.70} \\ \scriptsize{270}} & \makecell{\textbf{85.56} \\ \scriptsize{324}} & \makecell{\textbf{86.75} \\ \scriptsize{378}} & \makecell{\textbf{86.60} \\ \scriptsize{432}} & \makecell{\textbf{89.50} \\ \scriptsize{486}} & \makecell{\textbf{88.60} \\ \scriptsize{540}} & \makecell{\textbf{90.94} \\ \scriptsize{594}} & \makecell{\textbf{88.94} \\ \scriptsize{648}} & \makecell{\textbf{90.44} \\ \scriptsize{702}} & \makecell{\textbf{90.75} \\ \scriptsize{756}} & \makecell{\textbf{91.80} \\ \scriptsize{810}} \\ \hline
\textbf{40} & \makecell{\textbf{81.56} \\ \scriptsize{144}} & \makecell{\textbf{84.00} \\ \scriptsize{216}} & \makecell{\textbf{87.30} \\ \scriptsize{288}} & \makecell{\textbf{86.60} \\ \scriptsize{360}} & \makecell{\textbf{87.94} \\ \scriptsize{432}} & \makecell{\textbf{90.10} \\ \scriptsize{504}} & \makecell{\textbf{89.44} \\ \scriptsize{576}} & \makecell{\textbf{89.80} \\ \scriptsize{648}} & \makecell{\textbf{88.90} \\ \scriptsize{720}} & \makecell{\textbf{90.90} \\ \scriptsize{792}} & \makecell{\textbf{90.70} \\ \scriptsize{864}} & \makecell{\textbf{91.80} \\ \scriptsize{936}} & \makecell{\textbf{91.44} \\ \scriptsize{1008}} & \makecell{\textbf{91.90} \\ \scriptsize{1080}} \\ \hline
\textbf{50} & \makecell{\textbf{81.90} \\ \scriptsize{180}} & \makecell{\textbf{85.75} \\ \scriptsize{270}} & \makecell{\textbf{87.40} \\ \scriptsize{360}} & \makecell{\textbf{88.80} \\ \scriptsize{450}} & \makecell{\textbf{89.06} \\ \scriptsize{540}} & \makecell{\textbf{89.40} \\ \scriptsize{630}} & \makecell{\textbf{89.10} \\ \scriptsize{720}} & \makecell{\textbf{90.44} \\ \scriptsize{810}} & \makecell{\textbf{91.56} \\ \scriptsize{900}} & \makecell{\textbf{92.20} \\ \scriptsize{990}} & \makecell{\textbf{90.60} \\ \scriptsize{1080}} & \makecell{\textbf{92.75} \\ \scriptsize{1170}} & \makecell{\textbf{92.75} \\ \scriptsize{1260}} & \makecell{\textbf{93.40} \\ \scriptsize{1350}} \\ \hline
\textbf{60} & \makecell{\textbf{82.30} \\ \scriptsize{216}} & \makecell{\textbf{86.75} \\ \scriptsize{324}} & \makecell{\textbf{88.60} \\ \scriptsize{432}} & \makecell{\textbf{88.90} \\ \scriptsize{540}} & \makecell{\textbf{88.80} \\ \scriptsize{648}} & \makecell{\textbf{90.90} \\ \scriptsize{756}} & \makecell{\textbf{90.44} \\ \scriptsize{864}} & \makecell{\textbf{91.75} \\ \scriptsize{972}} & \makecell{\textbf{92.25} \\ \scriptsize{1080}} & \makecell{\textbf{92.94} \\ \scriptsize{1188}} & \makecell{\textbf{91.56} \\ \scriptsize{1296}} & \makecell{\textbf{92.50} \\ \scriptsize{1404}} & \makecell{\textbf{93.44} \\ \scriptsize{1512}} & \makecell{\textbf{93.44} \\ \scriptsize{1620}} \\ \hline
\textbf{70} & \makecell{\textbf{82.20} \\ \scriptsize{252}} & \makecell{\textbf{86.20} \\ \scriptsize{378}} & \makecell{\textbf{89.80} \\ \scriptsize{503}} & \makecell{\textbf{89.60} \\ \scriptsize{630}} & \makecell{\textbf{90.20} \\ \scriptsize{756}} & \makecell{\textbf{90.75} \\ \scriptsize{882}} & \makecell{\textbf{90.75} \\ \scriptsize{1007}} & \makecell{\textbf{91.60} \\ \scriptsize{1134}} & \makecell{\textbf{91.90} \\ \scriptsize{1260}} & \makecell{\textbf{92.25} \\ \scriptsize{1386}} & \makecell{\textbf{92.94} \\ \scriptsize{1512}} & \makecell{\textbf{93.25} \\ \scriptsize{1638}} & \makecell{\textbf{93.90} \\ \scriptsize{1764}} & \makecell{\textbf{94.70} \\ \scriptsize{1889}} \\ \hline
\textbf{80} & \makecell{\textbf{83.30} \\ \scriptsize{288}} & \makecell{\textbf{87.06} \\ \scriptsize{432}} & \makecell{\textbf{89.06} \\ \scriptsize{576}} & \makecell{\textbf{90.75} \\ \scriptsize{720}} & \makecell{\textbf{90.44} \\ \scriptsize{864}} & \makecell{\textbf{91.25} \\ \scriptsize{1008}} & \makecell{\textbf{91.40} \\ \scriptsize{1152}} & \makecell{\textbf{92.06} \\ \scriptsize{1296}} & \makecell{\textbf{92.06} \\ \scriptsize{1440}} & \makecell{\textbf{92.40} \\ \scriptsize{1584}} & \makecell{\textbf{92.56} \\ \scriptsize{1728}} & \makecell{\textbf{93.75} \\ \scriptsize{1872}} & \makecell{\textbf{94.20} \\ \scriptsize{2016}} & \makecell{\textbf{94.70} \\ \scriptsize{2160}} \\ \hline
\textbf{90} & \makecell{\textbf{84.25} \\ \scriptsize{324}} & \makecell{\textbf{87.60} \\ \scriptsize{486}} & \makecell{\textbf{90.06} \\ \scriptsize{648}} & \makecell{\textbf{90.40} \\ \scriptsize{810}} & \makecell{\textbf{90.60} \\ \scriptsize{972}} & \makecell{\textbf{91.44} \\ \scriptsize{1134}} & \makecell{\textbf{90.60} \\ \scriptsize{1296}} & \makecell{\textbf{92.06} \\ \scriptsize{1458}} & \makecell{\textbf{92.70} \\ \scriptsize{1620}} & \makecell{\textbf{93.10} \\ \scriptsize{1782}} & \makecell{\textbf{92.50} \\ \scriptsize{1944}} & \makecell{\textbf{94.20} \\ \scriptsize{2106}} & \makecell{\textbf{95.00} \\ \scriptsize{2268}} & \makecell{\textbf{94.80} \\ \scriptsize{2430}} \\ \hline
\end{tabular}
\vspace{5mm}
\caption{5-class classification accuracy values for different numbers of sampling points using the intra-latitudinal sampling scheme}
\label{tab3}
\end{table*}

\subsection{Two-dimensional Sampling Scheme}
We conducted an experiment to illustrate the importance of designing a suitable sampling scheme and how a different sampling scheme can affect the deficiency detection outcome. So far, the latitude-based sampling scheme has been employed to reduce sensitivity to rotations in the placement of nanosurfaces. 
We now aim to investigate how the sampling area is affected when the sampling scheme is modified from a latitude-based approach to a two-dimensional, point-based sampling along both latitude and longitude.
This means the algorithm will have more freedom in choosing sampling points on the surface of the ARS sphere.

\begin{figure}[hbt!]
\centerline{\includegraphics[width=\columnwidth]{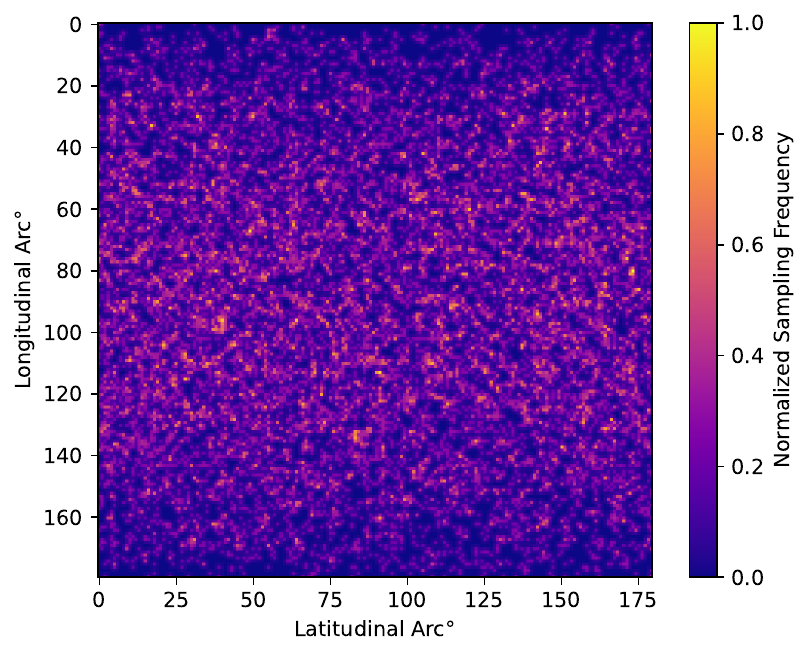}}
\caption{Heatmap of frequently sampled areas in 2D sampling scheme}
\label{fig:2D-heatmap}
\end{figure}

We modified the CL framework by changing PSO's search space.
The dimension of the search space corresponds to the desired number of sampling points and represents the number of optimization parameters.
The range of values in each dimension is between $0$ and $32400$ (ARS image size). The other parts of the framework, including the feature extraction, classification, and optimization loop, are kept the same. In the 2D sampling setup, the number of iterations increased to 1000 from 80, which was the case in latitude-based sampling, because of a more complex search space and an increase in the number of optimization parameters. The algorithm was run for 10 different runs for $1\%$ sampling rate, and the heatmap of frequently sampled areas is shown in Fig.~\ref{fig:2D-heatmap}. As can be seen, the sampled points are mostly concentrated in the middle area. This region corresponds to 2 sides of the sphere.

 The resulting sampling patterns, which mostly include random patterns and concentrated patterns on the sides of the scatterometry sphere, suggest that the optimization algorithm converges to a local minimum. Despite good overall deficiency detection accuracy, the results were not consistent across all runs. More importantly, concentrated sampling points on a specific side of the sphere translate to sensitivity to the orientation of nanosurface placement, which makes this sampling scheme less robust when applied to unseen data.
 The reason for this behavior can stem from the small size of the dataset.
 
Therefore, it is important to use the appropriate latitude-based sampling. This prior constraint can also help establish a universal sampling scheme for ARS in future investigations involving different surface types and deficiencies.

\section{Discussion}

\begin{figure}[hbt!]
\centerline{\includegraphics[width=\columnwidth]{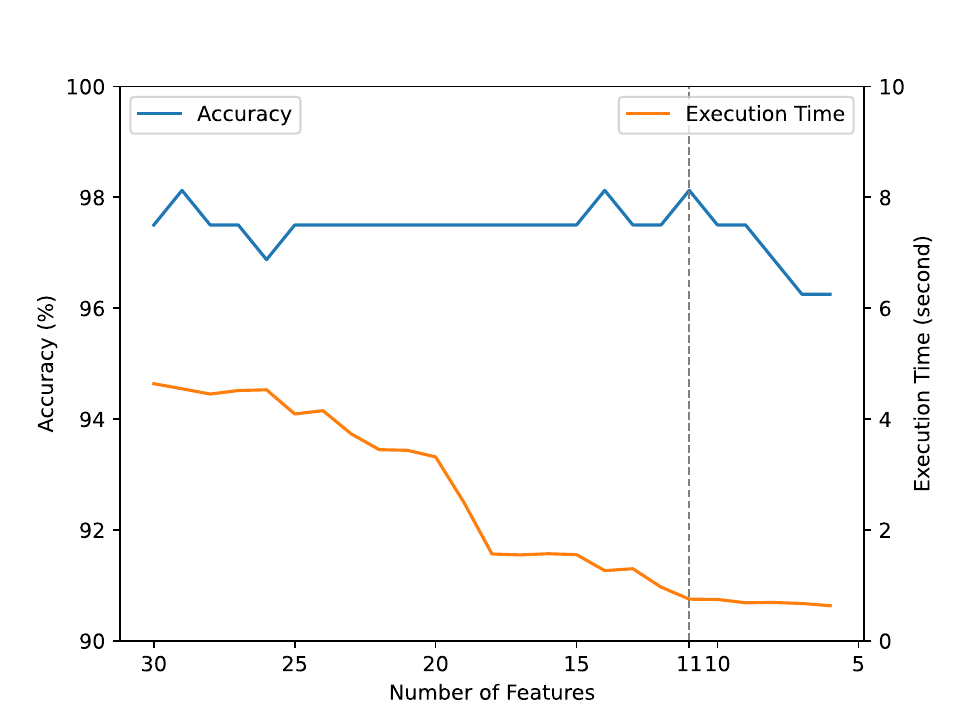}}
\caption{Deficiency detection accuracy and execution time of the feature extraction for different numbers of features in each iteration of RFE}
\label{fig6}
\end{figure}

The feature selection holds special importance in our study. In the optimization loop in the workflow of the proposed method, which can be seen in Fig.~\ref{fig1}, a feature extraction step is needed at each iteration, which includes extraction of each of the features from each of the images in the dataset. As a result, a larger set of features to extract will slow down the convergence process.
However, feature selection, while reducing the input space of the classifier, should not compromise its accuracy. Considering time and accuracy as two important factors, we run RFE and record the execution time and accuracy of the classifier to select the adequate number of features. These two factors are shown in Fig.~\ref{fig6}.
For different numbers of features. At each iteration of RFE, a single feature is eliminated, and the classification accuracy and execution time of the feature extraction are calculated.

As can be seen, the number of features decreases from 30 and accuracy follows a constant trend up to 11 features. This constant trend could be due to redundancy in the features, where some features are highly correlated with one another, and removing them does not significantly affect the classifier's performance. 
The classifier might be relying on a subset of features that already capture the most relevant information, making the additional features less important. 

With 11 features, we achieve considerably faster feature extraction while keeping accuracy at its highest level.
This results in a good trade-off between model performance and execution time.
The feature selection process made feature extraction for ARS images 6.2 times faster, reducing the extraction time from 4.64 seconds to 0.75 seconds for the entire dataset.

Classification model choice also plays a special role in our study. As discussed in Section \ref{3A}, LDA is preferred due to its higher accuracy. In addition to this, the simplicity and speed of LDA are also important factors in our work.
LDA can be trained significantly faster in comparison to other classification models like neural networks \cite{kumar2006artificial, bishop2006pattern}. This saves significant time in the optimization loop of the proposed framework because the classifier should be trained and tested at each iteration of the PSO algorithm. In addition, LDA can be trained on smaller datasets \cite{way2010effect}.
This characteristic is especially useful for our study, as the number of available data samples is limited. In contrast, many other machine learning models, particularly deep learning approaches, require large amounts of labeled data to achieve high performance \cite{lecun2015deep}.

To have a better understanding of the advantage of intra-latitudinal sampling, a closer look at the values in Table~\ref{tab3} is necessary.  For instance, for 14 latitudinal arcs and a 10\% sampling rate, the algorithm reaches 86.06\% accuracy using 252 sampling points. In contrast, with 2 latitudinal arcs and a 70\% sampling rate, which includes the same number of sampling points (252), the algorithm only reaches 82.20\% accuracy. This indicates that intra-latitudinal sampling increases accuracy by 3.86\% without any increase in the number of sampling points. In other words, a higher diversity of columns, even with lower sampling rates, achieves better accuracies compared to a lower number of columns with higher sampling rates.

In a real-world production setting, the proposed framework could significantly improve the efficiency of ZnO nanosurface inspection by reducing the number of required sampling points and accelerating ARS imaging. 
Traditional AFM or SEM methods are often slow, require complex sample preparation, and are not suitable for high-throughput inline industrial inspections. Our proposed framework enables faster defect detection while keeping high accuracy. For instance, in a ZnO thin-film production line,  manufacturers could integrate this algorithm to speed up inline inspection, allowing for rapid identification of surface defects, and roughness variations. ARS along with CL framework, helps minimize  production bottlenecks and makes sure only high-quality ZnO surfaces proceed to subsequent processing stages. 

Furthermore, the reduced sampling requirement has the potential to lower computational costs and hardware needs and make ARS imaging more practical for large-scale manufacturing. A possible hardware manufacturing can include a scatterometry sphere with mounted sensors in a fixed position instead of a single moving detector. In that case, reducing the density of sensors at different sampling arcs makes the manufacturing of these scatterometry spheres more practical.


Lastly, despite the promising results, this study has room for improvement. The current framework was trained and tested on a relatively small dataset of simulated ARS images. Due to the complexity of simulating such data, the dataset size is limited, which may contribute to the variations observed in the accuracy results shown in Fig.~\ref{fig3} and Table~\ref{tab3} when changing the number of sampling points. For instance, variations in accuracy occur when changing the number of sampled latitudinal arcs in data with Gaussian and salt-and-pepper noise, as seen in Fig.~\ref{fig3}.

While simulations provide a controlled environment for ARS data generation, future studies should also consider using experimental ARS data to validate the applicability of the framework to real-world scenarios.
In addition, studying scatterometry images of another dataset that includes images of a different nanosurface can validate the generalizability of the method over different nanosurface types.




\section{Conclusion}

This study presented a compressed learning framework designed for ZnO nanosurface deficiency detection using angle-resolved scatterometry data. Data simulations were carried out using discrete dipole approximation to generate the required data in a specific angle-resolved scatterometry setup.
The proposed framework successfully reduced the amount of required data for the detection of deficiency levels in ZnO nanosurface by sampling the latitudinal arcs that are most contributing.
The framework's robustness is demonstrated across various noise conditions, which shows its potential for real-world applications in nanoscale inspection. 
The results indicate that a significantly smaller set of sampling points can achieve nearly the same accuracy as full data sampling.
This work lays the groundwork for future studies in rapid and precise quality control in nanosurface manufacturing using scatterometry data.

\section{Acknowledgment}

This work was funded by the Deutsche Forschungsgemeinschaft (DFG, German Research Foundation) – 497286574

\bibliography{ref}
\bibliographystyle{IEEEtran}

\begin{IEEEbiography}[{\includegraphics[width=1in,height=1.25in,clip,keepaspectratio]{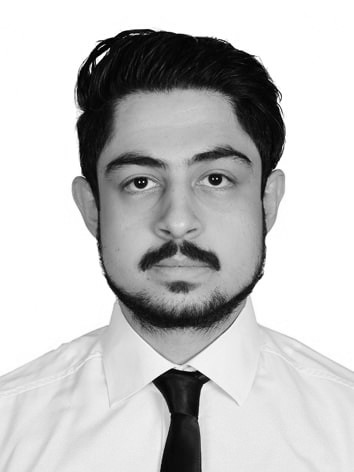}}]{Mehdi Abdollahpour} received the B.Sc. degree in Electrical Engineering (Electronics) and the M.Sc. degree in Biomedical Engineering (Bioelectric) from the University of Tabriz, Iran, in 2017 and 2019, respectively. He is currently pursuing a Ph.D. degree in Electrical Engineering at the University of Bremen, Germany. His research interests include compressed sensing, signal processing, image processing, and machine learning.
\end{IEEEbiography}

\begin{IEEEbiography}[{\includegraphics[width=1in,height=1.25in,clip, trim=6pt 0pt 6pt 0pt, keepaspectratio]{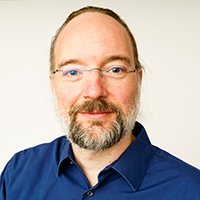}}]{Carsten Bockelmann}  received Dipl.-Ing. and Ph.D. degrees in electrical engineering from the University of Bremen, Germany, in 2006 and 2012, respectively. Since 2012, he has been a Senior Research Group Leader with the University of Bremen, coordinating research activities regarding the application of various frameworks like compressive sensing, DMD, Event-based sampling, and machine learning to communication problems. His research interests include 6G, massive machine-type communication, ultra-reliable low latency communications and industry 4.0 as well as health communications.
\end{IEEEbiography}

\begin{IEEEbiography}[{\includegraphics[width=1in,height=1.25in,clip,keepaspectratio]{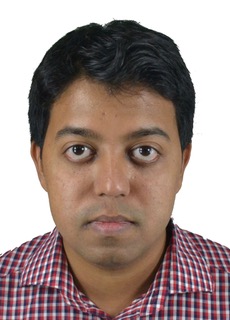}}]{TAJIM MD HASIBUR RAHMAN} received the B.Sc. degree in Aeronautical Engineering, with major in Aviation Electronics (Avionics) from Military Institute of Science and Technology (MIST) of Bangladesh in 2015 and the M.Sc. degree in Electrical Communication Engineering, with major in Opto-Electronics from the University of Kassel, Germany in 2022. He is currently pursuing a Ph.D. degree in Production Technology at the University of Bremen, Germany. His research interest includes computational photonics, microelectronics, functional materials, optical metrology and sensorics.
\end{IEEEbiography}

\begin{IEEEbiography}[{\includegraphics[width=1in,height=1.25in,clip,trim=0pt 5pt 0pt 0pt,keepaspectratio]{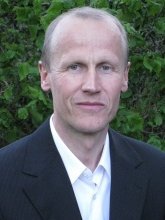}}]{Armin Dekorsy}   (IEEE Senior, 2018) is a professor at the University of Bremen, where he is the director of the Gauss-Olbers Space Technology Transfer Center and heads the Department of Communications Engineering. With over eleven years of industry experience, including distinguished research positions such as DMTS at Bell Labs and Research Coordinator Europe at Qualcomm, he has actively participated in more than 65 international research projects, with leadership roles in 17 of them. He is a Senior Member of the IEEE Communications and Signal Processing Society and a member of the VDE/ITG Expert Committee on Information and System Theory. He co-authored the textbook 'Nachrichtenübertragung, Release 6, Springer Vieweg,' which is a bestseller in the field of communication technologies in German-speaking countries. His research focuses on signal processing and wireless communications for 5G/6G, industrial radio, and 3D networks.
\end{IEEEbiography}

\begin{IEEEbiography}[{\includegraphics[width=1in,height=1.25in,clip,keepaspectratio]{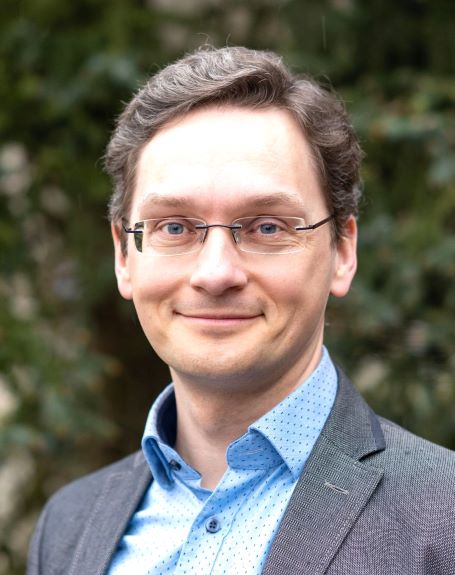}}]{ANDREAS FISCHER}
Andreas Fischer (M'16) studied electrical engineering, completed his PhD in the field of optical metrology at the Technische Universität Dresden in 2009 and his habilitation in 2014. Since 2016, he has been a Full Professor with the Production Engineering Department, University of Bremen, Germany, where he is currently the Director of the Bremen Institute for Metrology, Automation, and Quality Science (BIMAQ). As one of several awards, he received the Measurement Technique Prize 2010 of the Society of University Professors of Measurement Technique in Germany. In 2021, he further received an ERC Consolidaor Grant for enabling indirect optical geometry measurements. His research interests include optical measurement principles for production processes and flows, in-process applications of model-based measurement systems, and the investigation and overcoming of fundamental limits of measurability.
\end{IEEEbiography}

\EOD

\end{document}